\documentclass[layout=traditional, journal=jctcce, manuscript=article]{achemso}

\usepackage[version=3]{mhchem} 
\usepackage{xcolor}       

\author{Joshua Edzards}
\email{joshua.edzards@uni-oldenburg.de}
\affiliation{Carl von Ossietzky Universit\"at Oldenburg, Institute of Physics, 26129 Oldenburg, Germany}
\author{Julia Santana Andreo}
\affiliation{Carl von Ossietzky Universit\"at Oldenburg, Institute of Physics, 26129 Oldenburg, Germany}
\author{Holger-Dietrich Sa{\ss}nick}
\affiliation{Carl von Ossietzky Universit\"at Oldenburg, Institute of Physics, 26129 Oldenburg, Germany}
\author{Caterina Cocchi}
\email{caterina.cocchi@uni-oldenburg.de}
\affiliation{Carl von Ossietzky Universit\"at Oldenburg, Institute of Physics, 26129 Oldenburg, Germany}

\title{Benchmarking Selected Density Functionals and Dispersion Corrections for MOF-5 and its Derivatives}

\abbreviations{IR,NMR,UV}
\keywords{metal organic framework, density-functional theory, exchange–correction functional, dispersion correction}

\begin{document}

\begin{tocentry}

\includegraphics{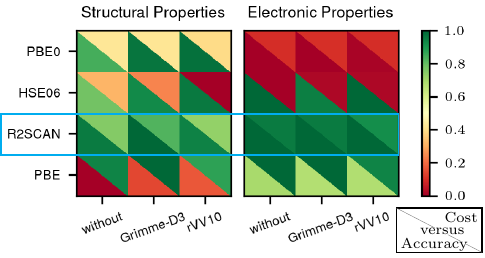}

\end{tocentry}


\begin{abstract}
Accurate computational predictions of metal-organic frameworks (MOFs) and their properties is crucial for discovering optimal compositions and applying them in relevant technological areas. This work benchmarks density functional theory (DFT) approaches, including semi-local, meta-GGA, and hybrid functionals with various dispersion corrections, on MOF-5 and three of its computationally predicted derivatives, analyzing structural, electronic, and vibrational properties. Our results underline the importance of explicitly treating van der Waals interactions for an accurate description of structural and vibrational properties, and indicate the meta-GGA functional R2SCAN as the best balance between accuracy and efficiency for characterizing the electronic structure of these systems, in view of future high-throughput screening studies on MOFs.
\end{abstract}

\newpage

\section{Introduction}

Metal-organic frameworks (MOFs) are an established class of materials formed by an organic network connected by metal nodes~\cite{kalm+18sa,guil-edda21acr,alle+21jacs,chen+22acr}.
Their structural flexibility and chemical versatility have brought them under the spotlight in crucial technological areas, including energy conversion and storage, as well as gas sensing and adsorption~\cite{sumi+12cr,guo+22am,gang+22rsca,kneb-caro22natn,ma+22aem,pal23mcf}. The current experimental efforts aimed at developing sustainable recipes for the synthesis of MOFs~\cite{juli+17gc,bhak+23nee} are mirrored by the implementation of efficient computational schemes that are suitable not only for simulating the fundamental properties of existing systems~\cite{ryde+19ats,mosl-tan22jpcl,wang+23jpcc,edza+23jpcc} but also for predicting new compositions and crystal structures~\cite{yu+23jpcl,vand+23npjcm,edza+24jcp,sass+24ic}.

High-throughput screening calculations based on first-principles methods have shown their potential in exploring the configurational space of technologically relevant material classes, ranging from semiconducting photocathodes~\cite{anto+21am,sass-cocc22jcp,sass-cocc24npjcm,schie+24ats,schi+24afm,sass-cocc25ats} to halide perovskites~\cite{naka-sawa17jpcl,jaco+19afm,li+20jcp,lee+24apr}, and from thermoelectrics~\cite{fan-orga21jmcc,hong+21jpcc,xion+22jacs} to complex oxides for batteries~\cite{wu-sun21acsml,yosh+22acsami}. These studies have contributed to expanding the range of available compounds for specific applications and paved the way for discovering new materials. The predictive power of \textit{ab initio} high-throughput screening approaches is, however, crucially influenced by the accuracy of the underlying approximations. In particular, the performance of density functional theory (DFT), the workhorse method for quantum-mechanical simulations in computational materials science, is extremely susceptible to the adopted exchange-correlation functional. While the extensive use of DFT has led to practical recipes and common knowledge for the most established materials~\cite{maur+19armr,burs+22acie}, extensive benchmarks are mandatory for hybrid systems~\cite{drax+14acr,hofm+21pccp} and new compounds~\cite{borl+19jctc,stol-lato20jpcc,sass-cocc21es}. 

In this work, we investigate the performance of different flavors of exchange-correlation functionals and dispersion correction schemes on the structural, electronic, and vibrational properties of MOF-5, a prototypical metal-organic framework~\cite{edda+00jacs}, and three of its variants with a Sr metal node in place of Zn and/or hydroxyl-functionalized organic linkers, identified as energetically stable in a recent high-throughput screening study based on DFT~\cite{edza+24jcp}. We examine how the generalized gradient approximation (GGA) in the Perdew-Burke-Ernzerhof (PBE) implementation~\cite{perd+96prl}, the metaGGA functional R2SCAN~\cite{furn+20jpcl}, and the hybrid functionals PBE0~\cite{adam+99jcp} and HSE06~\cite{hse03,hse06} predict lattice parameters, bond lengths, partial charges, electronic gaps, density of states, and phonon dispersions in MOF-5. Two schemes for dealing with van der Waals (vdW) interactions are considered, and their performance is contrasted against plain functionals, not including any treatment of dispersion corrections. The obtained results and a thorough analysis of the computational costs reveal that R2SCAN provides the optimal trade-off between accuracy and numerical efforts, and that accounting for vdW interactions is crucial to properly describe the structural and vibrational properties of these systems.

\section{Methods and Systems}
\subsection{Theoretical Background}
This work is based on DFT~\cite{hohn+64pr} in the Kohn–Sham (KS) framework~\cite{kohn+65pr}, where the effective single-particle Schr\"odinger equation
\begin{equation}
   \left[-\frac{1}{2}\nabla^2+v_{\text{eff}}(\Vec{r})\right]\phi(\Vec{r})=
   \epsilon_i^{\text{KS}}\phi_i(\Vec{r})
   \label{eq:KS}
\end{equation}
is solved for each (valence) electron in the system.
In Eq.~\eqref{eq:KS}, the KS eigenvalues $\epsilon_i^{\text{KS}}$ represent the energy per particle, while the eigenvectors $\phi_i(\Vec{r})$ denote the single-particle wave functions (KS states).
The Hamiltonian comprises the kinetic energy operator and the effective potential per particle $v_{\text{eff}} = v_{\text{ext}} + v_\text{H}+ v_{\text{xc}}$, where the external potential $v_{\text{ext}}$ describes the interaction between electrons and nuclei, the Hartree potential $v_\text{H}$ the repulsion experienced by each electron from the surrounding electron density, and the exchange-correction (xc) potential $v_{\text{xc}}$ all the remaining electron-electron interactions. Since the exact form of $v_{\text{xc}}$ is unknown, this term has to be approximated, and the adopted recipe ultimately determines the accuracy of the results.

Approximations for $v_{\text{xc}}$ are ranked according to the dependence of the xc energy on the density. 
The so-called local-density approximation introduced by Kohn and Sham~\cite{kohn+65pr} represents the lowest rung.
At the next level, we find the GGA, where $v_{\text{xc}}$ is determined from the local electron density and its gradient.
In this study, we utilize the PBE~\cite{perd+96prl} parameterization of GGA.
Further refinements of $v_{\text{xc}}$ incorporate the kinetic energy density, \textit{i.e.}, the second derivative of the density, leading to the so-called meta-GGA functionals.
Here, we adopt R2SCAN~\cite{furn+20jpcl}, a well-established representative of metaGGA functionals.
Hybrid functionals include a fraction of Hartree–Fock exchange to enhance the accuracy of the band gaps and mitigate the self-interaction error, which plagues all pure DFT functionals.
Notable examples of hybrid functionals are the \textit{global hybrid} PBE0~\cite{adam+99jcp}, incorporating 25\% of exact exchange, and the \textit{range-separated} hybrid functional HSE06~\cite{hse03,hse06}, including a short-range Hartree–Fock exchange component and a long-range component that aligns with the PBE functional. Both functionals are employed in this work.
Since the calculation of the Hartree-Fock exchange is particularly demanding, due to the nonlocal nature of the integrals involved, the application of an auxiliary density matrix method (ADMM) can reduce the costs without compromising on accuracy~\cite{guid+10jctc,merl+14jcp}.

Long-range and weak interactions, such as vdW forces, are not accurately captured by standard DFT functionals.
To address this limitation, an additional term is introduced into the effective potential. 
A widely adopted and numerically inexpensive method is the so-called Grimme-D3~\cite{grim+10jcp} correction, which empirically models dispersion forces and enhances the accuracy of DFT calculations for systems where such interactions are significant.
A more sophisticated recipe is provided by rVV10~\cite{saba+13prb}, where a non-local vdW correction is embedded into the xc functional.
Both approaches are considered in this work.

\subsection{Computational Details}
\label{sec:comput}

All calculations performed in this work are carried out with \texttt{CP2K}~\cite{kueh+20jcp} (version 2024.1) employing Goedecker-Teter-Hutter pseudopotentials.~\cite{goed+96prb}
A 2$\times$2$\times$2 k-mesh generated with the Monkhorst-Pack scheme~\cite{monk+76prb} is used in all runs in conjunction with cutoff and relative cutoff energies of 600~Ry and~100 Ry, respectively. 
The PBE, R2SCAN, PBE0, and HSE06 approximations for $v_{xc}$ are adopted in this work in conjunction with the Grimme-D3 and the rVV10 schemes for dispersion corrections, both adjusted to the corresponding functional. For Grimme-D3, the default settings for PBE are used also for PBE0, and HSE06, while for R2SCAN, the corresponding parameters are taken. For rVV10, the parameter $C = 0.0093$ remains fixed while $b$ is set to 10.0~\cite{peng+17prb,tran+19+prm} for PBE, PBE0, and HSE06, and to 11.95~\cite{ning+22prb} for R2SCAN.
The MOLOPT triple-zeta basis set with double polarization is applied for all simulations, except for the pre-optimization phase, where a double-zeta basis set with single polarization is utilized along with pressure and force thresholds of 200~bar and 0.005~Hartree/bohr, respectively.
For the volume optimization, these parameters are tightened to 100~bar and 0.0005~Hartree/bohr, respectively, while maintaining fixed angles to uphold the symmetry of the initial structure. To compute phonon frequencies, the cutoff and relative cutoff energies are increased to 2200~Ry and 500~Ry, respectively, while the pressure and force thresholds are reduced to 100~bar and 0.000019~Hartree/bohr, respectively.

The density of states (DOS) is computed in a 2$\times$2$\times$2 supercell to mimic a corresponding k-point sampling.
The partial charge analysis is conducted using the Bader scheme~\cite{bade+92cpl}, which has been successfully applied on MOFs~\cite{edza+23jpcc,sass+24ic,edza+24jcp}, using the package \texttt{Critic2}~\cite{oter+14cpc} (version 1.1) alongside the Yu-Trinkle integration technique.~\cite{yu+11jcp}
The harmonic interatomic force constants are calculated using \texttt{Phonopy}~\cite{togo23jpcm,togo23jpsj}, which applies finite displacements, here chosen with an amplitude of 0.01~\AA{}, to systematically perturb the atomic positions. To compute the vibrational response of the MOFs, 19 2$\times$2$\times$2 supercells, each containing 848 atoms, are generated from the primitive cell preserving its $Fm\overline{3}m$ cubic space group.
For the phonon calculations with the PBE0 functional, which are considerably more demanding than those with PBE and R2SCAN, unit cells (1$\times$1$\times$1) are used. 
The structures and plots are constructed with the Python package \texttt{aim2dat} \cite{aim2dat}.

Calculations are performed on infrastructure provided by the German National High Performance Computing Alliance (Emmy supercomputer in G\"ottingen) on Intel Sapphire Rapids Xeon Platinum 8468 double nodes, with a base clock frequency of 2.10~GHz, a maximum turbo clock frequency of 3.80~GHz, and 48 cores each. In the computational resources analysis, the number of core hours is calculated as $96 \cdot n \cdot t$, where 96 accounts for the total number of cores per node, $n$ is the total number of nodes, and $t$ the total time of the calculation in hours. 
\subsection{Construction of the Systems}

\begin{figure}
   \includegraphics{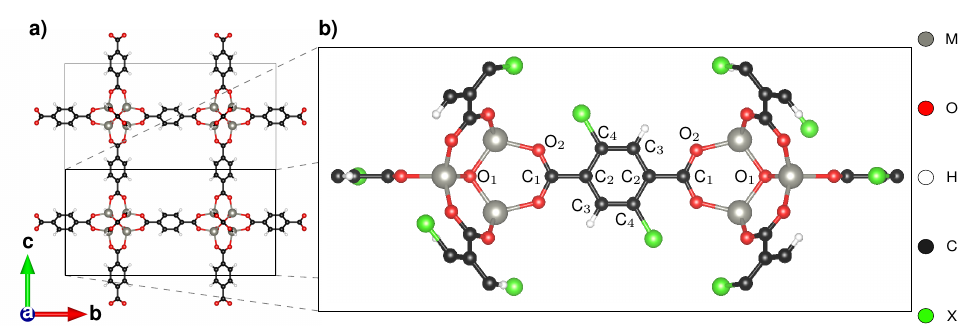}
   \caption{
  a) Conventional cubic cell of MOF-5 and b) zoom-in on the functionalized linker molecule in the coordinated framework with diagonal functionalization (X = H, OH). Zn and Sr are considered as metal nodes (M). The labels assigned to carbon and oxygen atoms are used in the analysis.}
   \label{fig:structure}
\end{figure}

The systems considered in this work are MOF-5 and derivatives in the structure proposed by Butler et al.~\cite{butl+14jacs, butl+23git}, characterized by a face-centered cubic lattice ($Fm\overline{3}m$ space group) with an experimental lattice constant of $a=25.87$~\AA{}~\cite{li+99nat}.
MOF-5 is characterized by Zn atoms interconnected via 1,4-benzene-dicarboxylate (BDC) linkers (Figure~\ref{fig:structure}a).
In earlier work, we computationally explored MOF-5 variants with different metal nodes and/or functionalized linker, identifying a set of energetically stable structures~\cite{edza+24jcp}.
Here, we consider four of such systems with Zn and Sr modes and with the BDC either H-passivated or functionalized with two OH groups in diagonal positions (Figure~\ref{fig:structure}b). 
The primitive cells with (without) OH-functionalization include 118 (106) atoms in total.

\section{Results}
\subsection{Structural Properties}
\begin{figure}
   \centering
   \includegraphics{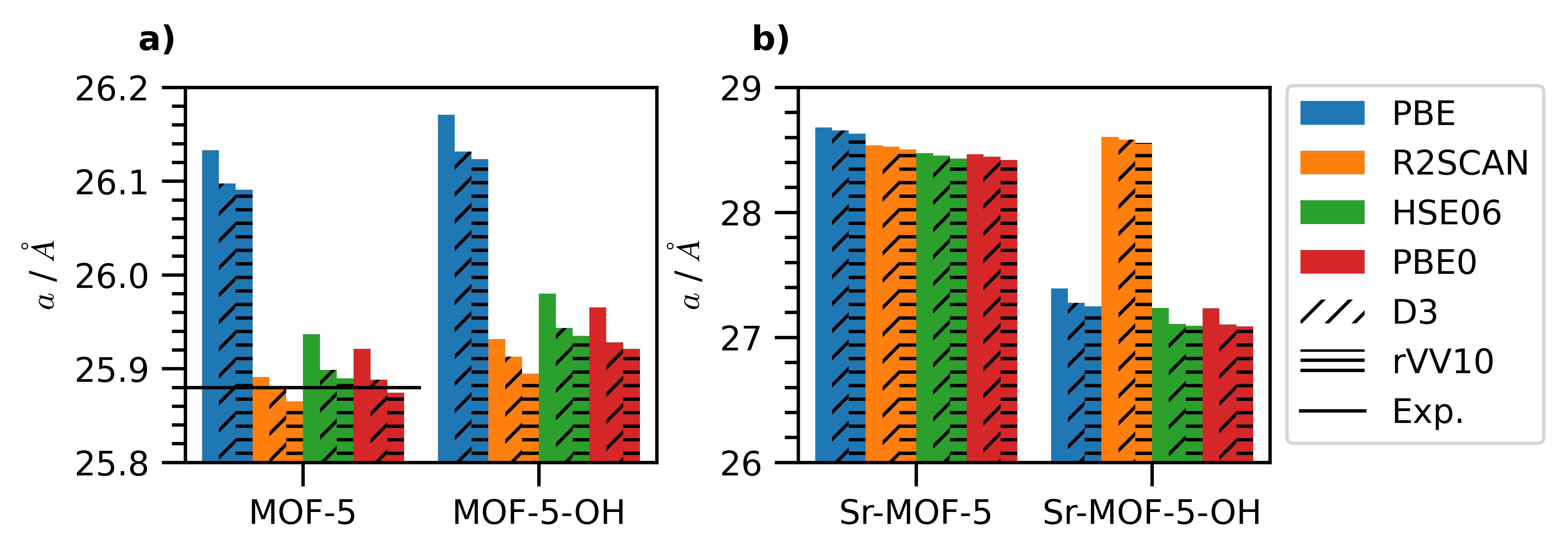}
   \caption{
  Lattice parameter $a$ of a) conventional MOF-5 and b) its Sr-substituted counterpart with H-passivated (left) and hydroxyl functionalized BDC (right). The experimental reference available for conventional MOF-5~\cite{li+99nat} is indicated by a horizontal bar.
  }
   \label{fig:lattice}
\end{figure}

We begin our analysis from the influence of exchange-correlation (xc) functionals and dispersion corrections on the structural properties of the MOFs, focusing initially on the lattice parameter $a$. For conventional MOF-5, the adopted $v_{\text{xc}}$ approximation strongly affects the crystal volume (Figure~\ref{fig:lattice}a). Due to its known underbinding behavior~\cite{zhan+18njp}, the PBE functional overestimates $a$. While the results are improved by incorporating dispersion corrections, the PBE lattice constant exceeds the experimental value~\cite{li+99nat} by more than 0.2~\AA{} independent of the treatment of vdW forces (Table~S1).
In contrast, hybrid functionals and R2SCAN, when combined with the Grimme-D3 correction, yield lattice parameters in almost perfect agreement with experiment~\cite{li+99nat} (Figure~\ref{fig:lattice}a, left).
Neglecting dispersion corrections slightly overestimates $a$ for both hybrid functionals and R2SCAN. In the latter, the difference between the Grimme-D3-corrected result and the uncorrected one is only 0.02~\AA{}. On the other hand, rVV10 leads to an underestimation of the lattice parameter which is most pronounced with R2SCAN and least so with HSE06. Similar trends are obtained with PBE0 (Figure~\ref{fig:lattice}a, left). 
The addition of hydroxyl groups to the BDC linker does not qualitatively change the crystal structure (Figure~\ref{fig:lattice}a, right). PBE notably overestimates the lattice constant compared to all other functionals tested. The hybrid functionals, HSE06 and PBE0, predict larger volumes than R2SCAN. Including dispersion corrections reduces $a$, with a more pronounced effect with rVV10 than with the Grimme-D3 method.

Similar trends are obtained for the Sr-based MOF-5 with H-passivation (Figure~\ref{fig:lattice}b, left). PBE yields a larger lattice parameter compared to the hybrid and meta-GGA functionals. Incorporating vdW interactions using the Grimme-D3 and the rVV10 methods systematically reduces $a$ across all tested functionals. The variations obtained with different $v_{\text{xc}}$ and vdW treatments are consistent with the results for pristine MOF-5, see Table~S1. Note the significantly larger scale of the $y$-axis in Figure~\ref{fig:lattice}b compared to Figure~\ref{fig:lattice}a.

The same analysis performed on the OH-functionalized Sr-based MOF-5 reveals qualitatively different results, see Figure~\ref{fig:lattice}b, right. R2SCAN predicts the largest lattice constant, which is reduced by approximately 1.1~\AA{} by adopting PBE (Table~S1). The two hybrid functionals, HSE06 and PBE0, yield nearly identical lattice constants, smaller than the PBE value. 
The trend for vdW corrections is consistent with the other compounds: including either the Grimme-D3 and rVV10 
treatment systematically reduces $a$ compared to the uncorrected case.
Since experimental data for this predicted structure are missing, we can only compare our computational results. As discussed in Ref.~\citenum{edza+24jcp}, hydroxyl functionalization induces a significant torsion (approximately 15${^\circ}$) in the BDC linker of the Sr-based MOF-5. In the resulting structure, the oxoclusters are significantly distorted, leading to a corresponding decrease in the unit cell volume compared to the H-passivated counterpart~\cite{edza+24jcp}. While all functionals capture this effect (Figure~\ref{fig:lattice}b), R2SCAN converges to a different structural minimum with a different cell size compared to PBE and its hybrid derivatives (HSE06 and PBE0).

\begin{figure*}
   \centering
   \includegraphics{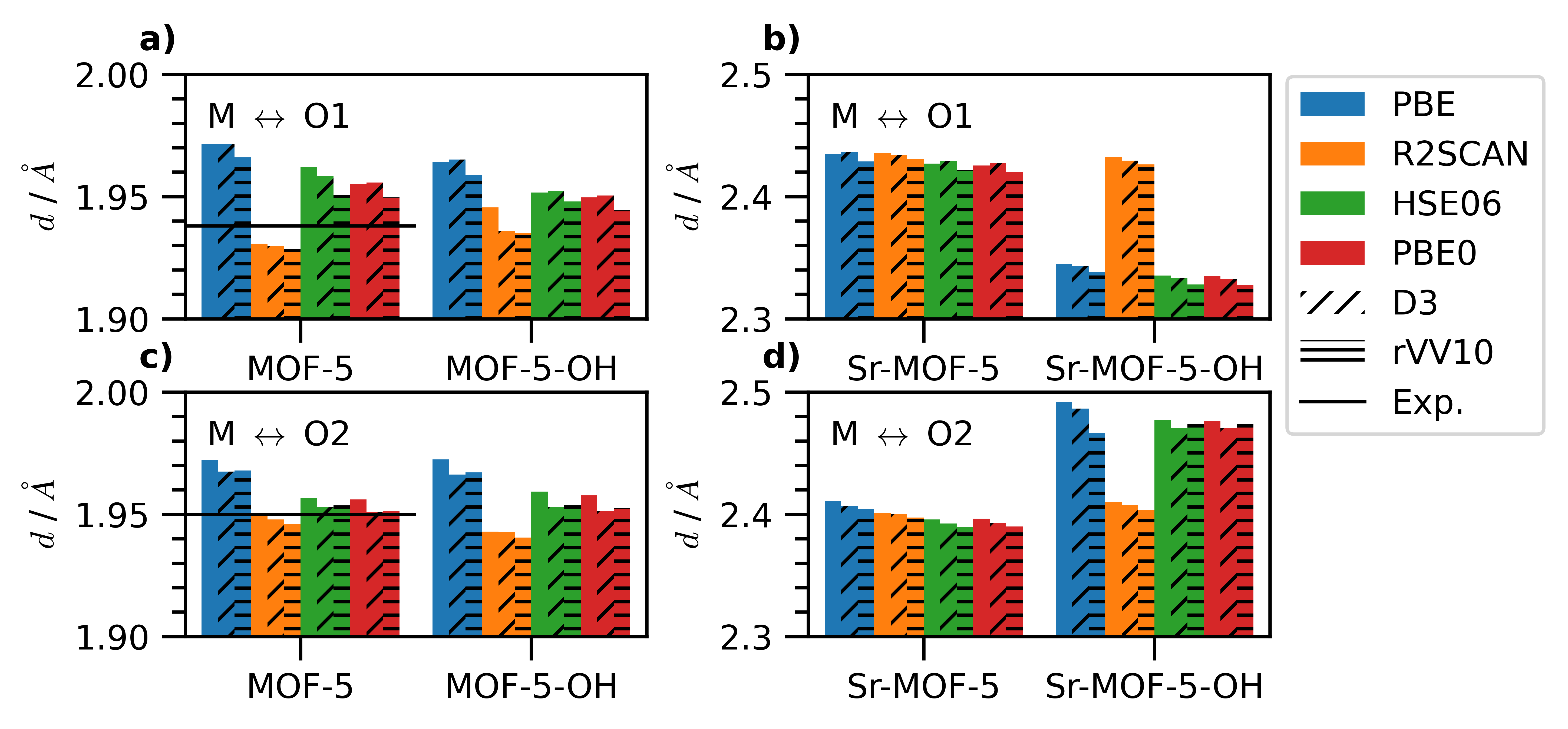}
   \caption{Interatomic distances between the metal atom and O1 (see Figure
   \ref{fig:structure}) in a) conventional MOF-5 and its OH-functionalized counterpart, as well as in the b) Sr-based variants with H- and OH-termination of the BDC linker. Interatomic distances between the metal atom and O2 (see Figure
   \ref{fig:structure}) in c) conventional MOF-5 and its OH-functionalized counterpart, as well as in the d) Sr-based variants with H- and OH-termination of the BDC linker.
  The experimental references available for conventional MOF-5~\cite{li+99nat} are marked by a horizontal bar.
  }
   \label{fig:dis_M_O}
\end{figure*}

To gain further insight into the impact of the xc functional on the structural properties of the analyzed MOF-5 variants, we examine key interatomic distances, specifically the separation between the metal node (M) and its neighboring oxygen atoms (O1 and O2, see Figure~\ref{fig:structure}b).
Starting with pristine MOF-5 (Figure~\ref{fig:dis_M_O}a, left), for which experimental data exists~\cite{li+99nat}, we notice trends consistent with those obtained for the lattice parameter (Figure~\ref{fig:lattice}). In particular, the PBE functional predicts the longest M-O distances. Incorporating dispersion corrections shortens these bonds, particularly with the rVV10 method. The Grimme-D3 scheme has a negligible effect on the M$\leftrightarrow$O1 distance and only a minor impact on M$\leftrightarrow$O2.
Hybrid functionals yield M$\leftrightarrow$O1 separations close to those from PBE (Figure~\ref{fig:dis_M_O}a, left), while M$\leftrightarrow$O2 distances are more similar to those from R2SCAN (Figure~\ref{fig:dis_M_O}c, left). The metaGGA approximation predicts the shortest bond lengths. Notably, the R2SCAN result (without dispersion correction) for the M$\leftrightarrow$O2 distance matches the experimental value~\cite{li+99nat}, as do PBE0+Grimme-D3 (Figure~\ref{fig:dis_M_O}c, left, and Table~S2). For the M$\leftrightarrow$O1 separation, R2SCAN (without dispersion correction) provides the best agreement with the experiment (Figure~\ref{fig:dis_M_O}a, left) with a discrepancy of less than 0.01~\AA{}. However, variations in bond lengths across different xc functionals and dispersion corrections are all within 0.03~\AA{} with respect to each other and the experimental value~\cite{li+99nat}.

The trends obtained for the xc functionals and vdW treatments in pristine MOF-5 are mirrored in its hydroxyl-functionalized counterpart.
PBE again predicts the largest interatomic distances, while R2SCAN yields the shortest. Hybrid functionals produce similar results, falling between those of PBE and R2SCAN (see Figure~\ref{fig:dis_M_O}a,c, right). However, a subtle difference emerges regarding dispersion corrections: the Grimme-D3 correction with PBE, HSE06, and PBE0 results in larger M$\leftrightarrow$O1 distances than the corresponding uncorrected functionals (Figure~\ref{fig:dis_M_O}a, right). The same is obtained with R2SCAN for M$\leftrightarrow$O2 (Figure~\ref{fig:dis_M_O}c, right). Consistent with the results for pristine MOF-5, the variations in interatomic distances due to different xc functionals and vdW treatments remain on the order of 0.01~\AA{}.

The predicted interatomic distances for the H-passivated Sr-based MOF-5 variant agree with the calculated lattice parameter. Figure~\ref{fig:dis_M_O}b and d (left panels) show the M$\leftrightarrow$O bond lengths. PBE predicts the largest values, followed by R2SCAN and the hybrid functionals. Similar to MOF-5, dispersion corrections influence the bond lengths. The Grimme-D3 method increases M$\leftrightarrow$O1 with all xc functionals except R2SCAN (Figure~\ref{fig:dis_M_O}b). Conversely, the longest Sr$\leftrightarrow$O2 distances are observed without dispersion corrections (Figure~\ref{fig:dis_M_O}d). Including the rVV10 correction yields the shortest metal-oxygen separations for all tested approximations for $v_{\text{xc}}$.

The structural analysis of OH-functionalized Sr-based MOF-5 reveals a more complex picture, although the calculated lattice parameters remain consistent with predictions from various xc functionals and vdW treatments. Unlike the other compounds examined so far, the Sr$\leftrightarrow$O1 distance is approximately 0.1~\AA{} shorter than the Sr$\leftrightarrow$O2 distance, due to the significant distortion of the BDC molecule induced by concomitant Sr-substitution and hydroxyl functionalization~\cite{edza+24jcp}. Consequently, the trends observed for Sr$\leftrightarrow$O1 (Figure~\ref{fig:dis_M_O}b, right) and Sr$\leftrightarrow$O2 (Figure~\ref{fig:dis_M_O}d, right) are mirrored.
The structures calculated using R2SCAN, which are less distorted than those predicted by the other functionals, result in similar distances with and without linker functionalization. Further details are provided in Tables~S1 and S2.
On the other hand, the PBE functional without dispersion correction exhibits the largest value for the distance with Sr$\leftrightarrow$O1.
However, adding the Grimme-D3 correction to PBE decreases Sr$\leftrightarrow$O1, in contrast with the effect obtained without it (Figure~\ref{fig:dis_M_O}b, right). This same trend is achieved for the hybrid functionals with Grimme-D3, although in these cases, the Sr$\leftrightarrow$O1 distance is shorter than that predicted by PBE. PBE0 and HSE06 yield nearly identical Sr$\leftrightarrow$O1 separation, see Table~S2. The rVV10 correction further decreases the distance. A similar pattern is obtained for Sr$\leftrightarrow$O2, where only the rVV10 correction in combination with the hybrid functionals increases the distance.

\subsection{Partial Charge Analysis}
\begin{figure}
   \centering
   \includegraphics{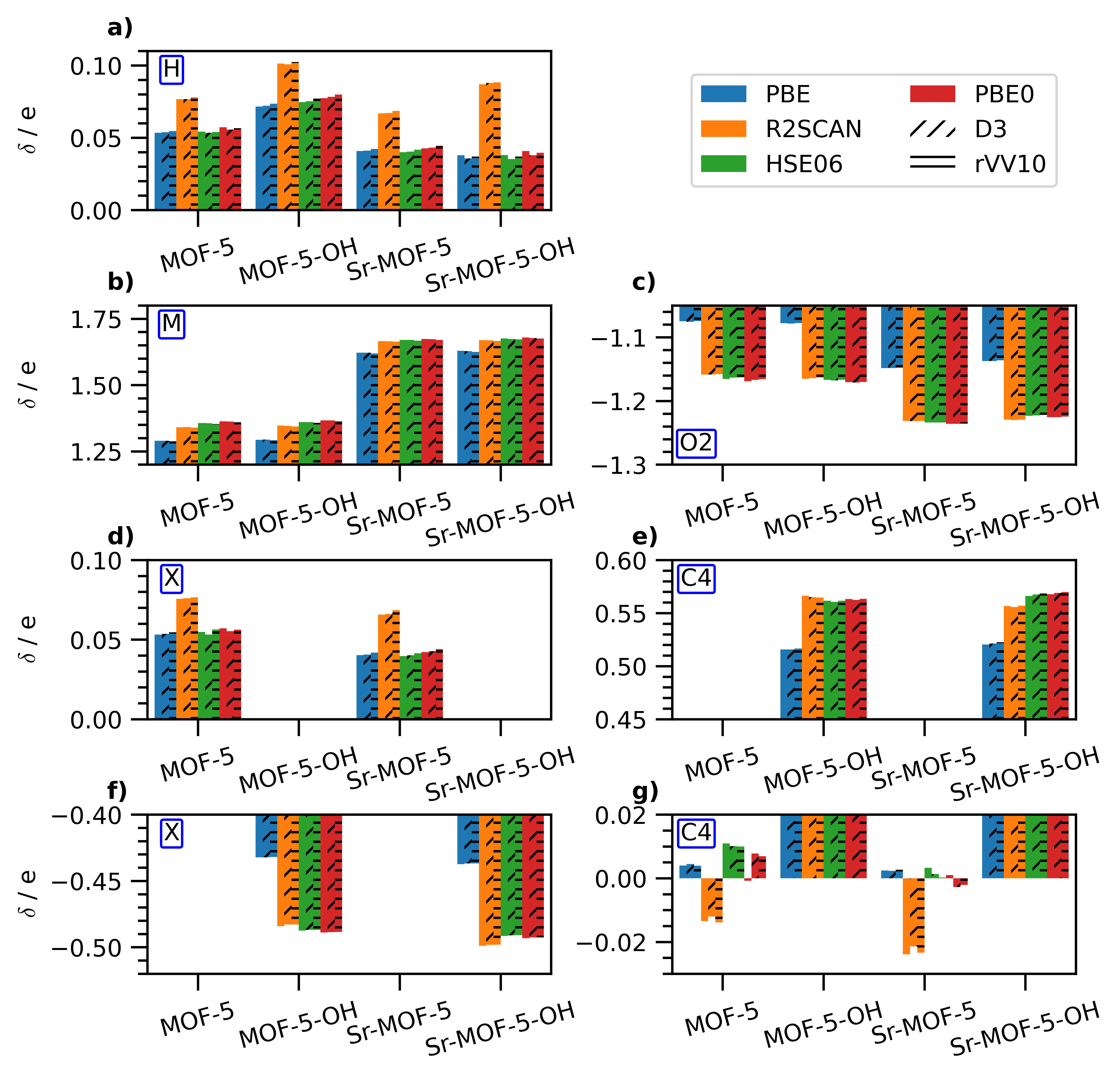}
   \caption{Partial charges computed with Bader scheme using different xc functionals and treatments of vdW interactions on a) H atoms bound to C3 in the BDC, b) on the metal node M, c) on O2, d) and f) on the functional group X (H and OH), and e) and g) on the C atoms to which they are attached. 
  }
   \label{fig:pc}
\end{figure}

We continue our analysis by examining the performance of the considered xc functionals and dispersion corrections on the atomic partial charges.
For the H atoms passivating C3 in the BDC linker (Figure~\ref{fig:structure}b), we notice similar trends with values between approximately 0.05~$e^-$ and 0.10~$e^-$ for all considered MOF-5 variants (Figure~\ref{fig:pc}a).
R2SCAN predicts the largest values followed by PBE0, HSE06, and PBE, which deliver almost identical results. The treatment of vdW forces plays a minor role here, being responsible for fluctuations one order of magnitude smaller than those produced by different $v_{\text{xc}}$ approximations (0.01~$e^-$). Nevertheless, the partial charges consistently increase going from no vdW treatment, to the Grimme-D3 scheme and the rVV10 correction. Further details are provided in Table~S4.

Different trends are obtained for the partial charges on the metal atoms, see Figure~\ref{fig:pc}b and Table~S5. Both Zn and Sr exhibit a more positive charge in MOF-5 than in their atomic phase. The hybrid functionals, PBE0 and HSE06, predict the largest partial charges, consistent with the inclusion of exact exchange which promotes charge localization. Conversely, the semi-local PBE functional yields the smallest partial charges for both metals, regardless of linker functionalization. The results obtained with R2SCAN are slightly lower but very close to those given by the hybrid functionals, suggesting that kinetic energy dependence in the metaGGA approximation mimics the influence of exact exchange in mitigating the limitations of semi-local functionals.
The choice of van der Waals correction has a negligible impact on the partial charges. Finally, we note that the consistently higher partial charges predicted for Sr compared to Zn are due to the $s$-character of the outermost shell of Sr compared to the completely filled $d$-shell of Zn, as discussed in previous work~\cite{edza+24jcp}. 

The oxygen atom O2 exhibits a significant negative charge accumulation (Figure~\ref{fig:pc}c), consistent with its high electronegativity and spatial proximity to the positively charged metal nodes. The magnitude of the partial charges on O2 in the Sr-based MOFs exceeds those on MOF-5, mirroring the trends for the metal nodes (compare Figure~\ref{fig:pc}b). Here, the divergence between the semi-local PBE functional and higher-level approximations is particularly pronounced, with PBE predicting absolute charge values approaching 0.1~$e^-$. While PBE0 predicts the most negative partial charges on O2 across all compounds, HSE06 and R2SCAN yield very similar results (see Table~S6 for details). This behavior reflects the localization of the electrons on oxygen atoms bound to metals, a textbook example often illustrated for metal oxides of the limitations of semi-local functionals and the need for more sophisticated treatments~\cite{fran+07prb,schr+10prb,kost-kagh24prb}. The adopted flavor for vdW corrections does not affect these findings.

We next examine the partial charges on the functional group of BDC. These results are summarized in Figure~\ref{fig:pc}d and Figure~\ref{fig:pc}f, where ``X" stands for H and OH in the protonated and hydroxyl-functionalized structures, respectively. In MOF-5 and its Sr-based counterpart, the partial charges on H are positive as expected, ranging between 0.05 and 0.1~$e^-$ depending on the chosen xc functional. The trends reported in Figure~\ref{fig:pc}d are qualitatively similar to those in Figure~\ref{fig:pc}a, both in terms of absolute values and $v_{\text{xc}}$ performance. R2SCAN yields the largest values followed by the hybrid functionals and PBE, which perform similarly (see Table~S7). The vdW treatment leads to small fluctuations for H, as discussed in Figure~\ref{fig:pc}a, with rVV10 predicting larger partial charges.

The partial charges on OH are negative as expected, and their absolute values vary within a few hundredth unit charges in MOF-5 and its Sr-substituted counterpart (Figure~\ref{fig:pc}f). The presence of oxygen in the functional group results in the semi-local functional delivering the least negative partial charges for both compounds by about 0.05~$e^-$ in magnitude compared to the higher-level functionals. R2SCAN, HSE06, and PBE0 produce very similar results, consistent with the trends discussed in Figure~\ref{fig:pc}c. The employed dispersion correction method does not significantly affect partial charges in the hydroxyl-functionalized MOF-5 (Figure~\ref{fig:pc}f).

We finally analyze the partial charges on C4, the carbon atom of the BDC molecule anchoring the functional group (see Figure~\ref{fig:structure}b and Table~S8). As expected, the partial charge values reflect the slightly electropositive character of the H termination and the strong electronegativity of the OH group across all the considered systems (Figure~\ref{fig:pc}e,g).
In the H-passivated MOF-5 variants, the partial charges on C4 are near zero, fluctuating around 0.01~e$^-$. For conventional MOF-5 (Zn metal node), R2SCAN predicts negative results. The magnitude of the partial charges is smallest with the Grimme-D3 scheme, followed by results obtained by neglecting vdW interactions, and largest when the dispersion corrections are included with rVV10. Conversely, PBE and hybrid functionals predict small positive charges -- on the order of 0.02~e$^-$ or less -- except for PBE0 without dispersion corrections, which yields negative charges on C4 of approximately -0.001e$^-$. In Sr-based MOF-5, the differences between the functionals are more remarkable. Again, R2SCAN predicts negative partial charges for C4, while the Grimme-D3 scheme yields the smallest absolute value of $\sim$0.02~e$^-$. A negative partial charge of $\sim$-0.003~e$^-$ is given by PBE0+D3 and PBE0+rVV10, while no treatment of vdW interactions leads to positive charges of the order of 0.001~e$^-$. A vanishing partial charge is predicted by HSE06 in conjunction with rVV10. Using the Grimme-D3 scheme or neglecting vdW interactions with this hybrid functional leads to small positive charges on C4. PBE yields positive charges of the order of 0.002~e$^-$ in tandem with all  considered vdW treatments.

In the hydroxyl-functionalized MOF-5 variants, the C4 atom exhibits positive partial charges around +0.55~e$^-$. The atomic charges, regardless of the xc functional and vdW treatment used, are similar regardless of the metal mode. PBE yields the smallest charges, consistent with its behavior on the OH group itself (Figure~\ref{fig:pc}f). Conversely, in the hydroxyl-functionalized MOF-5, R2SCAN predicts the largest charges, although the difference compared to the HSE06 and PBE0 results is only about 0.01~e$^-$. When Zn is exchanged with Sr, R2SCAN yields smaller values compared to the hybrid functionals PBE0 and HSE06, with the inclusion of dispersion corrections leading to small fluctuations (Figure~\ref{fig:pc}f).


\subsection{Electronic Properties}
\begin{figure}
   \centering
\includegraphics[width=0.8\textwidth]{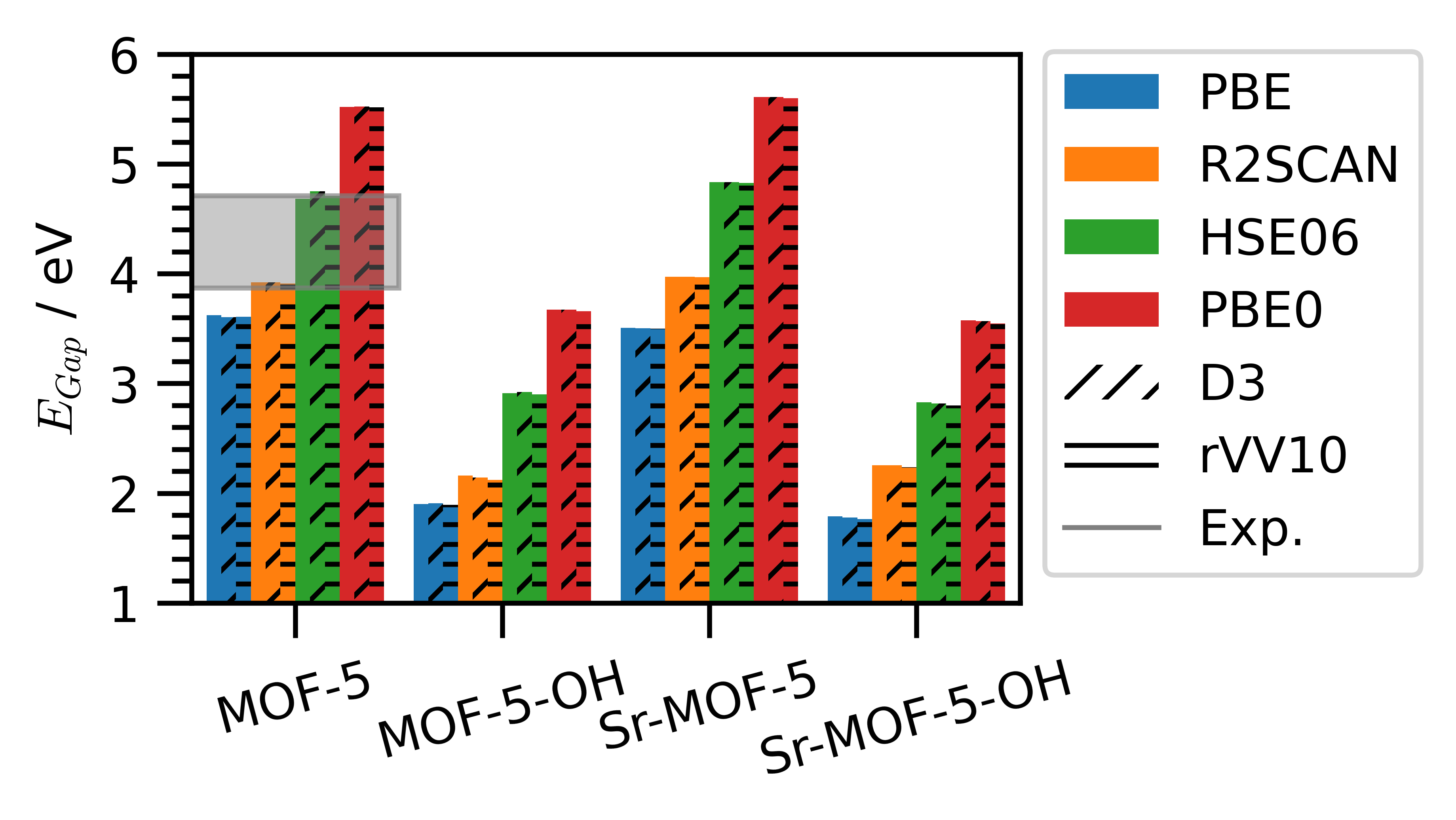}
   \caption{Fundamental gaps computed for conventional MOF-5, its Sr-substituted variant, and their OH-functionalized derivative using different xc functionals and vdW treatments. 
   The range of available experimental values for conventional MOF-5~\cite{hang+19jmatsci,chen+21rkmc,fabr+22acsml} is indicated by a gray shaded rectangle.
  }
   \label{fig:band_gap}
\end{figure}

We now inspect the electronic properties that are strongly influenced by the adopted xc functional, as expected.
As shown in Figure~\ref{fig:band_gap}(see also Table~S9), the band gap predicted by PBE for conventional MOF-5 is 3.6~eV while including a fraction of exact exchange increases it to 4.7~eV (HSE06) and 5.5~eV (PBE0). The result given by R2SCAN is between those~\cite{sass-cocc21es}, namely 3.9~eV. While the influence of vdW corrections on the band gap is two orders of magnitude smaller, careful inspection of Figure~\ref{fig:band_gap} reveals subtle differences. Excluding dispersion corrections or using the Grimme-D3 method results in slightly larger band gaps compared to the rVV10 scheme.

In the absence of experimental references for electronic band gaps based on photoemission or transport measurements, we validate our DFT results by contrasting them against optical absorption data obtained for conventional MOF-5. Challenges related to this comparison have been extensively discussed in the literature~\cite{fabr+22acsml,andr+24amt}. While DFT calculations notoriously neglect relevant physical effects, such as self-energy correction and electron-hole correlations, in MOFs, which combine a crystalline structure with organic building blocks, these effects largely cancel out~\cite{kshi+21jpcl}, in analogy to organic crystals~\cite{cocc-drax15prb}. Supported by this evidence, we can compare DFT band gaps with optical absorption measurements. Given the wide range of reported optical data for MOF-5~\cite{alva+07cej,gasc+08chemsuschem,hang+19jmatsci,chen+21rkmc,pand+22jpcc} and the various approaches available to process them~\cite{fabr+22acsml}, we take as a reference the range comprised between 3.87~eV, extracted from a recent experimental study~\cite{chen+21rkmc}, and 4.71~eV, reported in Ref.~\citenum{fabr+22acsml} as Gaussian fits of earlier optical absorption measurements~\cite{chen+21rkmc,hang+19jmatsci}.

As shown in Figure~\ref{fig:band_gap}, the R2SCAN and HSE06 results are within the experimental range, with R2SCAN accurately reproducing the lower bound and HSE06 the upper one. The established ability of range-separated hybrid functionals like HSE06 to accurately predict optical gaps of solids~\cite{sass-cocc21es,sb+22es}, including MOFs,~\cite{fabr+22acsml,butl+14acsami,dona+22jcp,orel24jpcc} is well documented. The good performance of R2SCAN is particularly significant given the growing popularity of metaGGA functionals in MOF research~\cite{mcca+24cm,conq+25jcc} and their demonstrated success in reproducing electronic gaps in technologically relevant materials~\cite{sass-cocc21es,tran+21jcp,swat+23jctc}. As expected, the semi-local PBE functional drastically underestimates the band gap. Conversely, PBE0 overestimates it, due to the fraction of Hartree-Fock exchange therein not effectively screened as in HSE06. For this reason, PBE0 is often favored for non-periodic systems but less appropriate for crystalline solids~\cite{garz-scus16jpcl}.

With the knowledge gained from this analysis of the band gap of MOF-5, we can continue with the inspection of its variants. While the trends obtained for pristine MOF-5, including the impact of dispersion corrections, are preserved regardless of the adopted xc functional (Figure~\ref{fig:band_gap}), hydroxyl functionalization significantly reduces the gap~\cite{edza+24jcp}. Specifically, the PBE gap is below 2~eV~\cite{edza+24jcp}, R2SCAN yields approximately 2.2~eV, HSE06 results are close to 3~eV, and PBE0 predicts gaps of 3.8~eV. Substituting Zn with Sr does not substantially alter the gap: R2SCAN predicts a small increase, while PBE, HSE06, and PBE0 a slight decrease.


\begin{figure}
   \centering
   \includegraphics{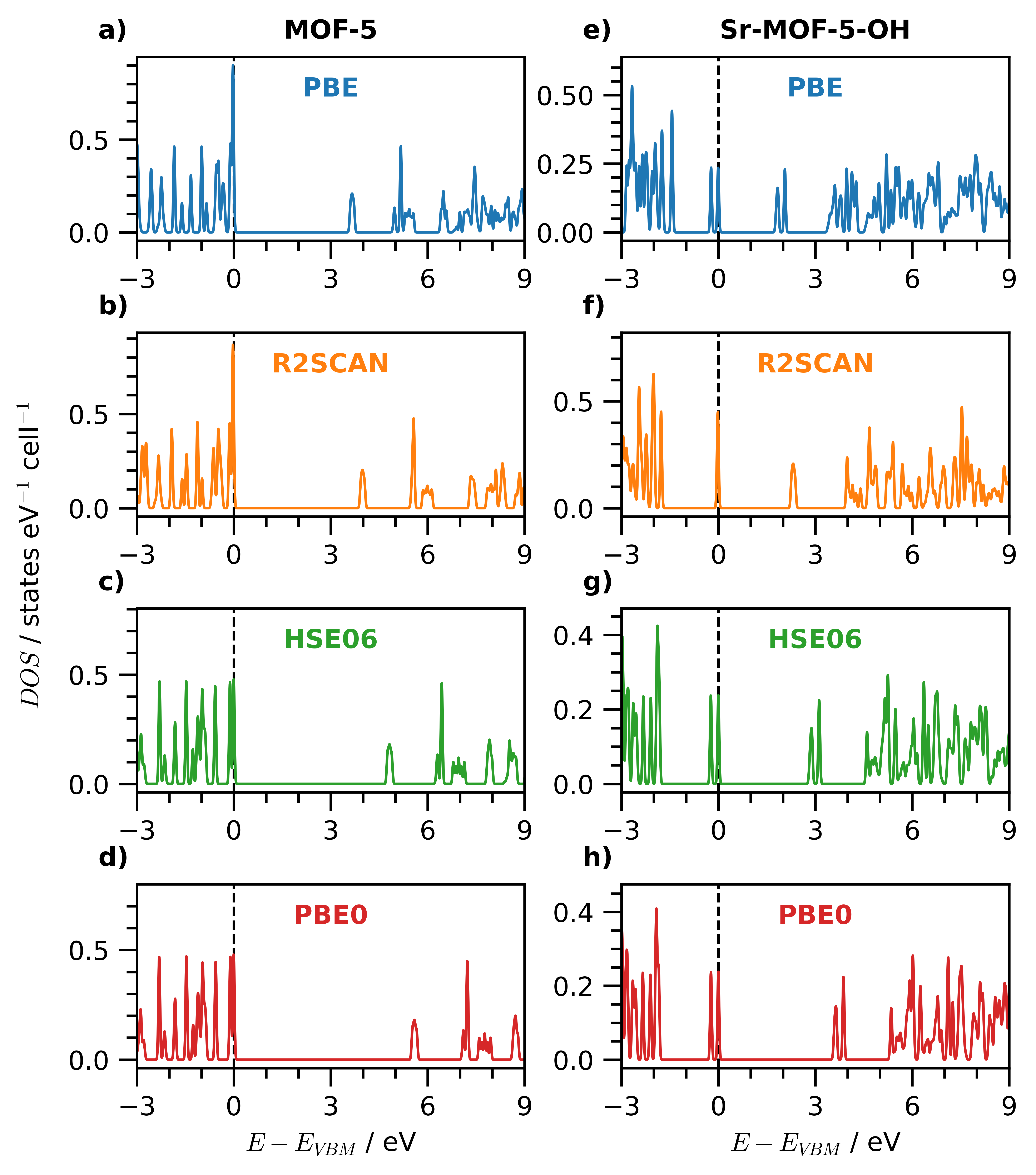}
   \caption{Density of states for conventional MOF-5 (left) using a) PBE, b) R2SCAN, c) HSE06, and d) PBE0, and its Sr-substituted hydroxyl functionalized counterpart (right) using e) PBE, f) R2SCAN, g) HSE06, and h) PBE0. In both cases, all functionals are supplemented with the Grimme-D3 method to deal with dispersion corrections. The energy scale is offset to the valence band maximum (VBM) set to zero. 
  }
   \label{fig:DOS}
\end{figure}

Next, we analyze the density of states (DOS) of conventional MOF-5 and its Sr-substituted OH-functionalized counterpart (Figure~\ref{fig:DOS}), focusing on how the different xc functionals reproduce the key features of the valence and conduction region. Given the minimal impact of different flavors of dispersion corrections, we present here only results obtained with the Grimme-D3 scheme. Further details are reported in Figures~S1-S3. 
For conventional MOF-5, all xc functionals reproduce almost identically the prominent peaks in the occupied and unoccupied regions, see Figure~\ref{fig:DOS}a-d. No significant differences between the plots are visible, apart from the absolute energies of the conduction states, since all DOS are aligned to the valence band maximum. Subtle variations in the relative peak energies can be attributed to the specific approximations employed for $v_{\text{xc}}$. For instance, in the unoccupied region, a pre-peak at approximately 5.5 eV is present in the DOS calculated with PBE, PBE0, and HSE06, likely due to the semi-local treatment of correlation shared by these three functionals, as it is absent in the R2SCAN results (Figure~\ref{fig:DOS}b). 

The electronic structure of MOF-5 remains substantially unaltered in its Sr-substituted OH-functionalized variant, although some discrepancies arise depending on the adopted xc functional (Figure~\ref{fig:DOS}e-h). The PBE results show a dominant, sharp peak in the valence region around -2.5~eV (Figure~\ref{fig:DOS}d). This peak shifts to lower energies with both HSE06 and PBE0 (Figures~\ref{fig:DOS}g,h) while it is not featured by R2SCAN (Figure~\ref{fig:DOS}f), which, on the other hand, yields a different description of the highest occupied valence states compared to the other xc functionals. The most striking variation is the presence of only one peak both at the top of the valence band and at the bottom of the conduction band, in contrast to PBE, HSE06 and PBE0 which predict two peaks in both cases. The origin of this splitting was attributed to structural distortions~\cite{edza+24jcp}, which promote hybridization between the functional group and BDC as predicted by PBE, HSE06, and PBE0 calculations.

\subsection{Phonon Dispersion}
In the last part of our analysis, we investigate the phonon band structure computed with PBE, R2SCAN, and PBE0 in conjunction with the Grimme-D3 and rVV10 schemes for dispersion corrections.
Due to the higher computational costs, we considered only one hybrid functional (PBE0) with looser settings compared to PBE and R2SCAN (see Section~\ref{sec:comput}). Furthermore, we focused only on MOF-5, as exploring BDC functionalization increases the complexity of the potential energy surface, thus dramatically complicating the search for the structural minimum of the material.

\begin{figure}
   \centering
    \includegraphics[width=\textwidth]{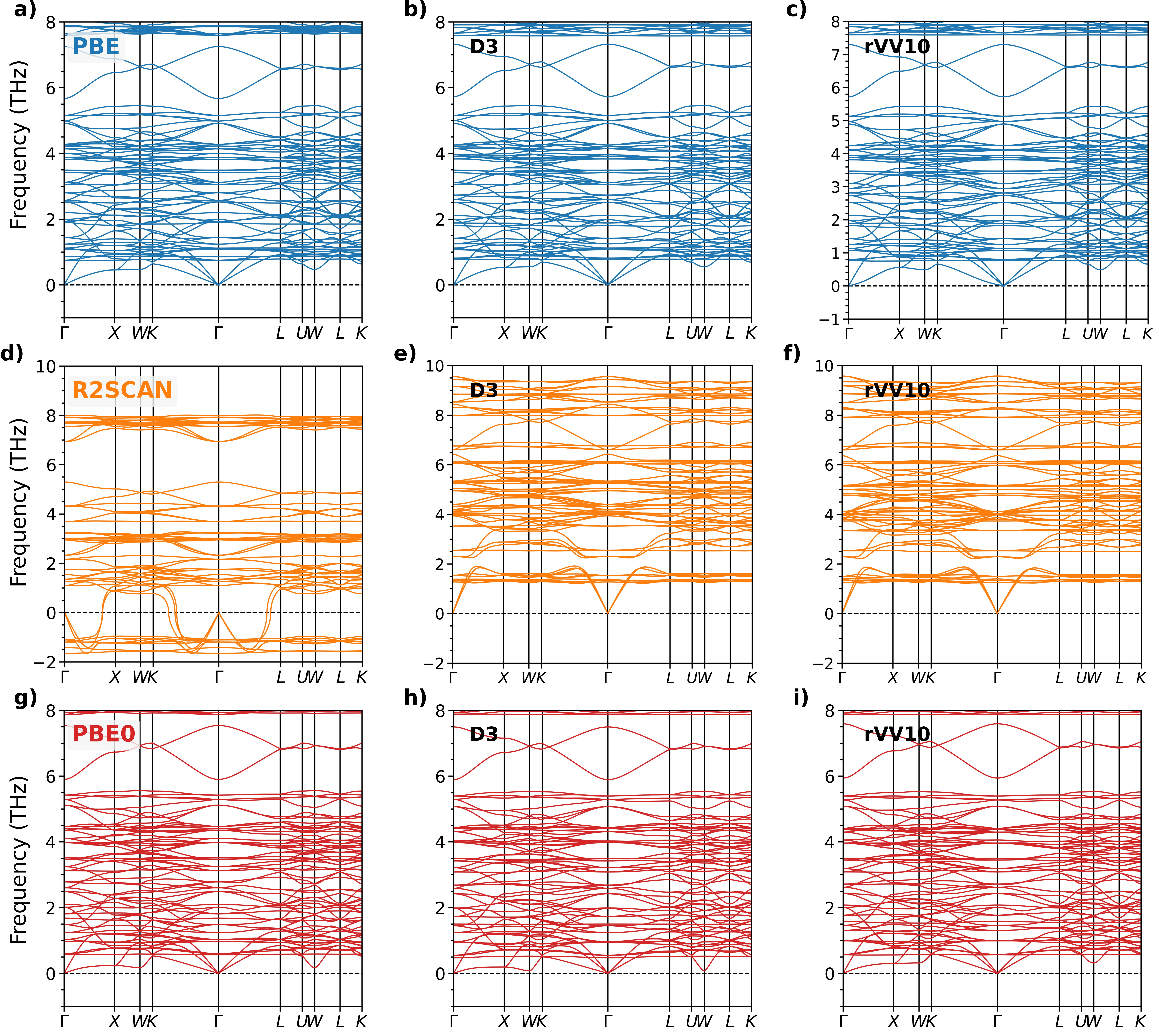}
   \caption{Phonon dispersion curves of primitive MOF-5 calculated with PBE a) without dispersion corrections, b) with the Grimme-D3 method, and c) with rVV10; with R2SCAN d) without dispersion corrections, e) with the Grimme-D3 method, and f) with rVV10; and with PBE0 g) without dispersion corrections, h) with the Grimme-D3 method, and i) with rVV10.}
   \label{fig:xc_phonons}
\end{figure}

No imaginary frequencies appear in the phonon band structure obtained with the PBE functional (Figure~\ref{fig:xc_phonons}a), indicating the dynamic stability of MOF-5 and the ability of this approximation to describe it. This finding is robust even in the absence of dispersion corrections and the inclusion of the Grimme-D3 or rVV10 induces only minor modifications to the phonon spectra that are hardly discernible from visual inspection of Figures~\ref{fig:xc_phonons}a-c. In contrast, calculations using R2SCAN without vdW corrections reveal imaginary frequencies (Figure~\ref{fig:xc_phonons}d), suggesting dynamic instability. This issue disappears by including dispersion corrections in both considered flavors (Figures~\ref{fig:xc_phonons}e,f), highlighting the role of vdW forces in stabilizing the structure predicted with this metaGGA functoinal. In general, R2SCAN+D3 and R2SCAN+rvv10 yield higher phonon frequencies compared to PBE, as visible from the larger frequency range visualized in Figures~\ref{fig:xc_phonons}d-f and by the qualitative differences in the band structures compared to Figures~\ref{fig:xc_phonons}a-c. The hardening of the material, as predicted by R2SCAN, is attributed to the smaller predicted lattice parameter (Figure~\ref{fig:lattice}): Since phonon frequencies are directly related to bond strengths, where tighter interatomic interactions lead to higher frequencies, a smaller lattice constant results in stronger interatomic forces and, consequently, stiffer phonon modes. Moreover, the R2SCAN bands appear noticeably flatter, particularly in the acoustic region, indicating reduced slope (i.e., lower group velocity) and suggesting lower phonon dispersion despite the enhancement of the local force constants.

The hybrid functional PBE0 yields phonon spectra generally similar to those given by PBE. MOF-5 is predicted to be dynamically stable regardless of the treatment of vdW interactions, see Figures~\ref{fig:xc_phonons}g-i. However, the phonon frequencies calculated with PBE0 are softer compared to those obtained with PBE and R2SCAN. This result is related to two combined factors: (i) the use of a smaller (1$\times$1$\times$1) supercell in the PBE0 calculations, which truncates long-range interatomic interactions, especially affecting low-frequency acoustic modes, and (ii) the inclusion of 25\% exact (Hartree–Fock) exchange in PBE0 which alters the interatomic force constants, thus affecting bond strengths and vibrational properties~\cite{reil+13tjocp}.

\begin{figure}[H]
   \centering
   \includegraphics[width=\textwidth]{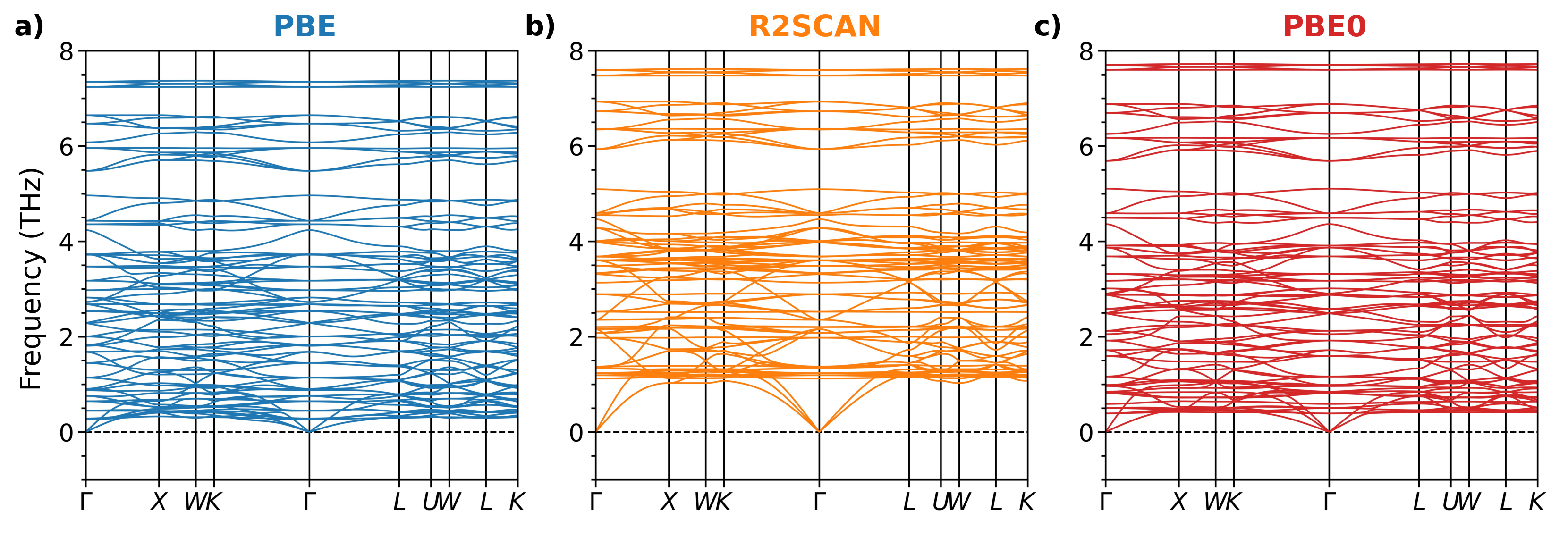}
   \caption{Phonon dispersion curves of Sr‐substituted MOF‐5 computed with a) PBE, b) R2SCAN, and c) PBE0 all supplemented by the Grimme-D3 method for treating vdW interactions.}
   \label{fig:xc_phonons_Sr}
\end{figure}

The trends obtained for MOF-5 are confirmed for its Sr-substituted counterpart, see Figure~\ref{fig:xc_phonons_Sr}. In this case, we adopted PBE, R2SCAN, and PBE0 supplemented only by the Grimme-D3 scheme for vdW corrections. Sr-substituted MOF-5 is predicted to be dynamically stable by all considered functionals, which is a relevant result for stimulating the synthesis of this compound. Similar to pristine MOF‑5, R2SCAN predicts a harder material compared to PBE and PBE0 (compare Figure~\ref{fig:xc_phonons_Sr} against Figure~\ref{fig:xc_phonons_Sr}a and c). In turn, PBE0 predicts higher phonon frequencies compared to PBE. Comparing Figure~\ref{fig:xc_phonons_Sr} with Figure~\ref{fig:xc_phonons}, we notice that the substitution of Zn with Sr leads to an overall softening of the phonon modes, which can be primarily ascribed to the higher atomic mass of Sr compared to Zn. 

From these results, it is evident that R2SCAN is the most suitable method to calculate phonon dispersion in MOF-5 and its variants. It corrects the over-softening featured by PBE and PBE0, likely due to its more accurate treatment of intermediate- and long-range interactions, as demonstrated in previous studies~\cite{sant+24jpm, zhan+17prb, sun+16nc, shao+23prb}. Furthermore, it is mandatory to include dispersion corrections to achieve a dynamically stable structure, although the differences between using D3 and rVV10 are not substantial.

\section{Discussion}
\begin{figure}
   \centering
   \includegraphics{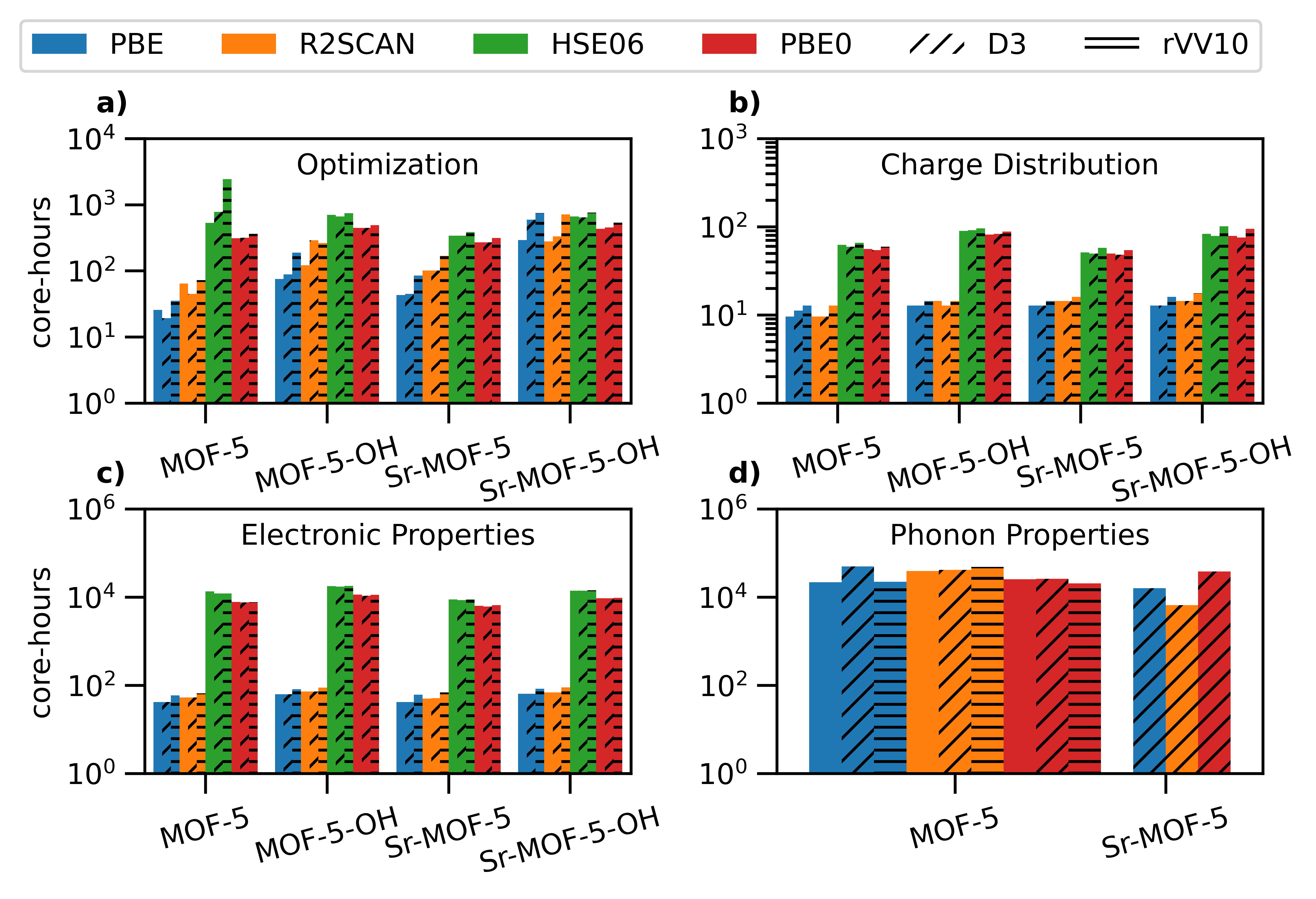}
   \caption{
   Core-hours required for a) structural optimization, b) partial charges, c) electronic properties, and d) phonon properties of MOF-5 and derivatives using different xc functionals and dispersion corrections. 
}
   \label{fig:core}
\end{figure}

The results presented in this study offer valuable guidance for choosing the most appropriate xc functional and treatment of dispersion corrections for MOF-5 and its derivatives, considering both accuracy and computational costs (Figure~\ref{fig:core}). For lattice parameters, R2SCAN+D3 provides excellent predictions (Figure~\ref{fig:lattice}) with similar trends obtained for interatomic distances (Figure~\ref{fig:dis_M_O}). While hybrid functionals HSE06 and PBE0, both with Grimme-D3, achieve comparable accuracy for H-terminated MOF-5 variants, they are significantly more computationally expensive, requiring approximately ten times the runtime (Figure~\ref{fig:core}a and Table~S10). Optimizing hydroxyl-functionalized structures is generally more demanding. For OH-terminated MOF-5 (Zn metal node), R2SCAN+D3 remains the optimal trade-off between accuracy and cost, though the advantage over hybrid functionals is reduced to approximately a factor of 5, see Figure~\ref{fig:core}a. For the hydroxyl-functionalized Sr-substituted MOF-5, the hybrid functionals are as expensive as R2SCAN and PBE, which, in turn, require almost 10 times more core hours than for the H-passivated counterpart. We can attribute this behavior to the considerable distortions in this system caused by the larger atomic radius of Sr compared to Zn and the presence of the OH group, which distorts the organic scaffold. These combined effects make structural optimization considerably more demanding than for the other systems examined in this work. We have to point out that the structures predicted with the hybrid functionals use the results from PBE as input. 

For the partial charges, HSE06+rVV10 provided the best agreement between our predictions and experimental references for conventional MOF-5 (Figure~\ref{fig:pc}). However, this method is computationally expensive, see Figure~\ref{fig:core}b and Table~S11. Since different xc functionals and dispersion corrections affect partial charges only by hundredths of an electron, the R2SCAN+D3 method is the most viable alternative for both Zn-based MOF-5 (with and without OH termination) and its Sr-substituted variants.

In contrast, electronic-structure calculations of MOFs are very sensitive to the choice of the xc functional, both in terms of accuracy and computational costs. As illustrated in Figure~\ref{fig:core}c (see also Table~S12), hybrid functionals require two orders of magnitude more core hours than PBE or R2SCAN, regardless of the system. Similarly, the rVV10 method for treating vdW interactions is significantly more expansive than Grimme-D3. Considering these factors, and the band gap results presented in Figure~\ref{fig:band_gap}, R2SCAN+D3 offers again the best balance between accuracy and computational efficiency.

Finally, the calculation of phonon dispersions is very expensive even with the PBE functional, see Figure~\ref{fig:core}d and Table~S13. The inclusion of vdW corrections, especially with the Grimme-D3 method, further increments the computational costs. Those imposed by R2SCAN are a factor 2 larger than those of PBE, with the specific vdW treatment not significantly affecting the trend. By examining Figure~\ref{fig:core}d, one could assume that the PBE0 requires less computing time than PBE and R2SCAN. However, PBE0 calculations were performed on a unit cell with 106 atoms, whereas PBE and R2SCAN calculations were conducted on 2$\times$2$\times$2 supercells with 848 atoms. This underscores the substantial increase in computational cost associated with hybrid functionals, despite yielding no significant improvement in the vibrational properties for conventional MOF-5 (Figure~\ref{fig:xc_phonons}). Remarkably for Sr-substituted MOF-5, calculations with R2SCAN+D3 are cheaper than those with PBE+D3 and PBE0+D3, see Figure~\ref{fig:core}d. This confirms that R2SCAN+D3 is the most suitable functional for calculating the phonon properties of MOF-5 and its variants.

\section{Summary and Conclusions}
In summary, we benchmarked the performance of four exchange-correlation functionals (PBE, R2SCAN, HSE06, and PBE0) and two dispersion correction schemes (Grimme-D3 and rVV10) on the structural, electronic, and vibrational properties of MOF-5 and its Sr-substituted and/or hydroxyl-functionalized derivatives using DFT. R2SCAN+D3 emerged as the optimal choice for calculating structural properties, accurately predicting lattice parameters and interatomic distances and requiring about ten times lower runtime than the hybrid functionals. Regarding partial charges, while HSE06+rVV10 provided the best agreement with experimental data for MOF-5, its high computational cost makes R2SCAN+D3 a cheaper alternative for comparable accuracy. Electronic structure calculations proved to be highly sensitive to the approximation of $v_{xc}$, as expected. Hybrid functionals demand two orders of magnitude more computational resources than PBE or R2SCAN, without a commensurate improvement in accuracy. Similarly, the rVV10 dispersion correction is significantly more expensive than the Grimme-D3 scheme. Considering these factors, R2SCAN+D3 provides again the best compromise between accuracy and computational efficiency. Phonon calculations, even with PBE, are computationally demanding, with the inclusion of dispersion corrections further increasing their costs. While R2SCAN calculations are more expensive than PBE by a factor of 2, the choice of the vdW treatment does not significantly affect this trend. Importantly, for the Sr-substituted MOF-5, R2SCAN+D3 is surprisingly less expensive than PBE+D3. This finding, along with the accuracy observed for other properties, confirms R2SCAN+D3 as the most suitable functional for calculating phonon properties in MOF-5 and its variants.

In conclusion, our work demonstrates that R2SCAN+D3 offers an excellent balance between accuracy and computational cost for predicting the intrinsic properties of MOF-5 and its derivatives. Our results provide valuable guidance for future computational studies on MOFs and related materials, enabling efficient and reliable predictions of key properties in high-throughput calculations, where more advanced calculations, based for example on many-body perturbation theory~\cite{kshi+21jpcl}, are prohibitive. As such, our study paves the way for accelerated computational design of MOFs with enhanced properties for applications in gas sensing and storage, which require simultaneously accurate predictions of structural, electronic, and vibrational properties.

\begin{acknowledgement}
This work was financed by the German Federal Ministry of Education and Research (Professorinnenprogramm III) and by the State of Lower Saxony (Professorinnen f\"ur Niedersachsen). J.E. appreciates financial support from the Nagelschneider Stiftung and J.S.A. acknowledges funding from the Evonik Stiftung. Computational resources were provided by the North-German Supercomputing Alliance (NHR), project nic00084.
\end{acknowledgement}

\begin{suppinfo}
Additional information regarding structural and electronic properties and computational costs is provided in the Supporting Information.

\end{suppinfo}

\providecommand{\latin}[1]{#1}
\makeatletter
\providecommand{\doi}
  {\begingroup\let\do\@makeother\dospecials
  \catcode`\{=1 \catcode`\}=2 \doi@aux}
\providecommand{\doi@aux}[1]{\endgroup\texttt{#1}}
\makeatother
\providecommand*\mcitethebibliography{\thebibliography}
\csname @ifundefined\endcsname{endmcitethebibliography}
  {\let\endmcitethebibliography\endthebibliography}{}

\end{document}